\documentclass[a4paper,11pt]{article}
\usepackage[hmargin=26mm,top=24mm,bottom=26mm]{geometry}
\usepackage[T1]{fontenc}
\usepackage{lmodern}
\usepackage{amsmath,amssymb,amsthm}
\usepackage{microtype}
\usepackage[ruled,vlined,linesnumbered]{algorithm2e}
\usepackage[hidelinks]{hyperref}

\DontPrintSemicolon
\SetAlCapNameFnt{\small\bfseries}
\SetAlCapFnt{\small}
\newtheorem{theorem}{Theorem}
\newtheorem{corollary}[theorem]{Corollary}
\newtheorem{lemma}[theorem]{Lemma}
\newtheorem{proposition}[theorem]{Proposition}

\title{A Tight 2-Approximation for Min-Power Edge Multicover}
\author{Amir Belgi\\
  {\small Independent Researcher, Israel}\\
  {\small\href{mailto:belgiamir251@gmail.com}{\texttt{belgiamir251@gmail.com}}}
  \and
  Zeev Nutov\\
  {\small The Open University of Israel, Ra'anana, Israel}\\
  {\small\href{mailto:nutov@openu.ac.il}{\texttt{nutov@openu.ac.il}}}}
\date{}
\hypersetup{
  pdftitle={A Tight 2-Approximation for Min-Power Edge Multicover},
  pdfauthor={Amir Belgi and Zeev Nutov},
  pdfkeywords={minimum power, edge multicover, activation, approximation algorithms, hardness of approximation}
}

\newcommand{\OPT}{\operatorname{OPT}}
\newcommand{\subs}{\subseteq}
\newcommand{\problem}[3]{\begin{center}\fbox{\begin{minipage}{\dimexpr\linewidth-2\fboxsep-2\fboxrule\relax}\underline{#1}\par\smallskip{\em Input:} #2\par\smallskip{\em Output:} #3\end{minipage}}\end{center}}

\begin{document}
\maketitle

\begin{abstract}
In {\sc Min-Power Edge Multicover} we seek a subgraph of minimum total
vertex power subject to lower bounds on the vertex degrees. The power
of a vertex is the maximum cost of a chosen incident edge. We give a
combinatorial 2-approximation. For maximum requirement $k$ its ratio is
at most $2-1/(4(k+1))$. The previous bound was $\min\{O(\log k),k+0.2785\}$.
A reduction from {\sc Vertex Cover} rules out every fixed ratio below two
under the Unique Games Conjecture and every fixed ratio below $\sqrt{2}$
unless $\mathrm P=\mathrm{NP}$, even on connected simple graphs.
For asymmetric endpoint costs with maximum endpoint ratio $\theta$,
we give a combinatorial $2\theta$-approximation and an
$O(\log(2+\min\{\theta,n\}))$-approximation, where $n$ is the number of
vertices.
\end{abstract}

\section{Introduction}\label{sec:introduction}
Let $G=(V,E)$ be an undirected graph with nonnegative edge costs $c_e$.
The {\bf power} of a vertex $v$ in an edge set $J\subs E$ is
\[
 p_J(v)=\max\{c_e:e\in J\cap\delta(v)\},
\]
with $p_J(v)=0$ if $v$ is isolated in $(V,J)$. The power of $J$ is
$p_J(V)=\sum_{v\in V}p_J(v)$. This objective arises in wireless networks,
where a station can reach several neighbors at the same transmission power.
Early work on minimum-power connectivity includes Chen and
Huang~\cite{ChenHuang1989} and Kirousis et al.~\cite{KirousisKranakisKrizancPelc2000}.
We study the following degree problem.

\problem{{\sc Min-Power Edge Multicover}}
{An undirected graph $G=(V,E)$, nonnegative rational costs $c_e$, and
integer requirements $r_v\geq0$.}
{An edge set $J\subs E$ of minimum power such that every vertex $v$ has
at least $r_v$ neighbors in $(V,J)$.}

Equivalently, choose a {\bf power assignment} $p=(p_v:v\in V)$ minimizing
$\sum_vp_v$. An edge $vu$ is {\bf active} if
$c_{vu}\leq\min\{p_v,p_u\}$, and each vertex needs at least $r_v$ active
neighbors. We write $n=|V|$, $k=\max_vr_v$, and $\OPT$ for the optimum
value.

The ratio improved from $O(\log^4 n)$~\cite{HajiaghayiKortsarzMirrokniNutov2007}
to $O(\log n)$~\cite{KortsarzMirrokniNutovTsanko2011}. Cohen and
Nutov~\cite{CohenNutov2015} obtained $O(\log k)$ for general costs and
a constant ratio for uniform costs. In 2011 they asked whether a constant
ratio is possible for general costs~\cite[Section~4 of the preprint]{CohenNutov2015}.

\begin{theorem}\label{thm:power}
{\sc Min-Power Edge Multicover} admits a combinatorial
2-approximation using $O(n+|E|\Delta^2\log(\Delta+1))$ arithmetic
operations, where $\Delta$ is the maximum degree. Its ratio is at most
$2-1/(4(k+1))$ for maximum requirement $k$.
\end{theorem}

In each round a vertex requests neighbors it can reach. The price rises
if a requested neighbor cannot reach back. We keep the upper half of each
vertex's powers. If every neighbor can reach back at least as often as it
is requested, the mean shortage at each vertex is at most $1/2$, whatever
its requirement (Lemma~\ref{lem:shortage}). The prices cover excess requests.
The saving from discarding the lower halves pays for the remaining repairs.

We show that the factor two is tight under the Unique Games Conjecture
by a reduction from {\sc Vertex Cover}.

\begin{theorem}\label{thm:lower}
For fixed $\gamma\geq1$ and $\eta>0$, a polynomial-time
$\gamma$-approximation for {\sc Min-Power Edge Multicover} gives a
polynomial-time $(1+\eta)\gamma$-approximation for {\sc Vertex Cover}.
The reduction uses connected simple graphs with positive integer
edge costs.
\end{theorem}

Taking $\eta$ arbitrarily small transfers the two {\sc Vertex Cover}
hardness bounds. The first is due to Khot and Regev~\cite{KhotRegev2008},
and the second follows from the $2$-to-$2$ Games
results~\cite{DKKMS2021,DKKMS2025,KhotMinzerSafra2023} and the
reduction in~\cite[Corollary~1.9]{KhotMinzerSafra2025}.

\begin{corollary}\label{cor:hardness}
For every fixed $\epsilon>0$, a polynomial-time
$(2-\epsilon)$-approximation for {\sc Min-Power Edge Multicover} would
refute the Unique Games Conjecture,
and a polynomial-time $(\sqrt{2}-\epsilon)$-approximation would imply
$\mathrm P=\mathrm{NP}$, even on the instances of Theorem~\ref{thm:lower}.
\end{corollary}

The hardness applies to unbounded $k$, while Theorem~\ref{thm:power}
gives a ratio below two for each fixed $k$.
For small $k$, Cohen and Nutov~\cite{CohenNutov2015} gave ratio $k+1/2$,
improved to $k+0.2785$ by Nutov, Huri, and Kortsarz~\cite{NutovHuriKortsarz2026}.
For {\sc Min-Power Edge Cover} ($k=1$), Kortsarz and Nutov~\cite{KortsarzNutov2009}
gave ratio $3/2$, C{\u a}linescu, Kortsarz, and Nutov~\cite{CalinescuKortsarzNutov2019}
obtained about $1.406$ in expectation, and Nutov, Kortsarz, and
Shalom~\cite{NutovKortsarzShalom2019} gave a ratio below $1.2785$.

By Hajiaghayi et al.~\cite[Theorem~3(i)]{HajiaghayiKortsarzMirrokniNutov2007},
a $\rho$-approximation for undirected {\sc Min-Cost $k$-Connected Subgraph}
now gives a $(2+2\rho)$-approximation for {\sc Min-Power $k$-Connected
Subgraph}, hence a constant ratio whenever $\rho$ is constant.
For undirected {\sc Min-Power $k$-Outconnected Subgraph}, where a given
root needs $k$ internally disjoint paths to each other vertex, the
reduction of Lando and Nutov~\cite{LandoNutov2010} gives ratio $2+4=6$.

\paragraph{Asymmetric endpoint costs.}
Panigrahi~\cite{Panigrahi2011} introduced the activation model, in which
the availability of an edge is a monotone function of the powers at its
endpoints. Nutov~\cite{Nutov2013} studied survivable connectivity in this
model. In the simpler formulation introduced by Nutov~\cite{NutovActivationSurvey},
an edge $e=vu$ is active when $p_v\geq c_e^v$ and $p_u\geq c_e^u$.
On finite power domains, one parallel edge per minimal activating pair
represents any monotone rule for $vu$. These edges give $v$ one neighbor.

The {\bf slope} $\theta$ is the least $\theta\geq1$ such that
$c_e^v\leq\theta c_e^u$ and $c_e^u\leq\theta c_e^v$ for every edge,
and $\theta=\infty$ if one threshold is zero and the other is positive.
Two zero thresholds impose no restriction.
For {\sc Activation Edge Cover} ($k=1$), Nutov, Kortsarz, and
Shalom~\cite{NutovKortsarzShalom2019} gave a bound in terms of a
vertex-based slope that is at most $\theta$ (Section~\ref{sec:asymmetric}).
For {\sc Activation Edge Multicover}, Nutov, Huri, and
Kortsarz~\cite{NutovHuriKortsarz2026} gave ratio
$O(\log k+\log\min\{\theta,n\})$.

For finite $\theta$, set $c_e=\max\{c_e^v,c_e^u\}$ on each edge.
For every edge set $J$, this increases the power at each vertex by a
factor of at most $\theta$. Theorem~\ref{thm:power} on the symmetric
instance therefore gives a combinatorial $2\theta$-approximation.

The next bound removes the $\log k$ term. For $\theta\geq n$ both
logarithmic bounds are $O(\log n)$.
\begin{theorem}\label{thm:activation}
{\sc Activation Edge Multicover} admits a polynomial-time
$O(\log(2+\min\{\theta,n\}))$-approximation.
\end{theorem}

\section{A combinatorial 2-approximation}\label{sec:analysis}
Keep the cheapest edge between each pair of neighbors. We assume
$r_v\leq\deg(v)$ for every $v$, since otherwise the instance is infeasible. Set isolated vertices aside at power zero.
Let $w_v$ be the $r_v$-th cheapest incident cost, or zero if $r_v=0$,
and put $W=\sum_v w_v$. Write $B_v=\{0\}\cup\{c_{vu}:u\in N(v)\}$.
Every feasible assignment has $p_v\geq w_v$, and we may restrict $p_v$ to $B_v$.

For any $p\geq w$, let $D_v(p)=(r_v-\deg_p(v))_+$ be the
\emph{shortage} at $v$, where $\deg_p(v)$ counts active neighbors and
$x_+=\max\{x,0\}$. We use the following bound on the cost after repair.
\begin{lemma}\label{lem:repair}
Powers $p\geq w$ can be extended to a feasible assignment of power at most
$\Psi(p)=\sum_v\bigl(p_v+w_vD_v(p)\bigr)$.
\end{lemma}
\begin{proof}
Visit each vertex once. If $v$ still lacks $D$ neighbors, at least $D$
of its $r_v$ cheapest edges are inactive. Vertex $v$ already affords
these edges. Raise the powers of $D$ corresponding neighbors to their
edge costs. This costs at most $Dw_v$, and powers only increase.
\end{proof}

\subsection{The algorithm}
In each round, vertex $v$ chooses a power $b_v\in B_v$ and requests
$r_v$ neighbors whose edges it can afford. Write $S_v$ for this set.
For an ordered pair $vu$, vertex $v$ pays a price $\lambda_{vu}$ to
request $u$, and $u$ earns the same price whenever it affords $vu$.
Thus $v$ chooses $(b_v,S_v)$ to minimize
\begin{equation}\label{eq:priced-star}
 b_v+\sum_{u\in S_v}\lambda_{vu}
       -\sum_{u:\,c_{vu}\leq b_v}\lambda_{uv}.
\end{equation}
For a given power take the $r_v$ lowest-priced affordable neighbors.
Prices rise for requests without an offer and fall for offers without
a request. 

Let $\Delta$ be the maximum degree. Return zero powers if $\Delta=0$. Otherwise run $T=16\Delta^2$ rounds. A counter $j_{vu}$ sets the
price $\lambda_{vu}=w_vj_{vu}/(8\Delta)$. Break ties in input order. In the selection step the average
is over all ways to take one entry from each current list, with
repetitions counted. Lemma~\ref{lem:conditional} gives its computation.

\begin{algorithm}[ht]
\caption{A combinatorial algorithm for {\sc Min-Power Edge Multicover}}\label{alg:power}
Set $j_{vu}\gets 4\Delta$ for both orders of each edge\;
\For{$t=1,\ldots,T$}{
 At each $v$, minimize~\eqref{eq:priced-star} at the current prices and append $b_v$ to its list\;
 For each $vu$, set $j_{vu}\gets\max\{0,j_{vu}+\mathbf{1}_{u\in S_v}-\mathbf{1}_{b_u\geq c_{vu}}\}$\;
}
Keep the largest half of each list\;
\ForEach{vertex $v$}{
 Replace its list by one entry minimizing the average of $\Psi$ over the resulting lists\;
}
Repair the shortages as in Lemma~\ref{lem:repair} and return the powers\;
\end{algorithm}

Record each power by incrementing its count in the sorted list $B_v$.
A round takes $O(|E|\log(\Delta+1))$ arithmetic operations after isolated vertices are removed.
At each vertex scan the neighbors by increasing edge cost, processing
ties together. Maintain the sum of incoming prices and the $r_v$
smallest eligible outgoing prices in a heap. After choosing the best
power, recover its requests in a second scan.

\subsection{The price bound}
Let $C$ be the mean total power of a round. Write $a_{vu}$ for the
fraction of rounds in which $v$ requests $u$, and $x_u(c)$ for the
fraction in which $u$ buys power at least $c$. The total excess of
requests over offers at $v$, counting each neighbor separately, is
\[
 e_v=\sum_{u\in N(v)}\bigl(a_{vu}-x_u(c_{vu})\bigr)_+.
\]
\begin{lemma}\label{lem:prices}
The lists produced by the pricing rounds satisfy
$C+\sum_v w_ve_v\leq\OPT+W/8$.
\end{lemma}
\begin{proof}
Write $g_{vu}=\mathbf{1}_{u\in S_v}-\mathbf{1}_{b_u\geq c_{vu}}$.
In an optimum solution each vertex can request $r_v$ active neighbors.
Its imbalances $g^*_{vu}$ are all nonpositive. Thus its priced objectives
sum to at most $\OPT$. Summing the local minima~\eqref{eq:priced-star}
gives, in each round,
\begin{equation}\label{eq:round}
 \sum_v b_v+\sum_{vu}\lambda_{vu}g_{vu}\leq\OPT,
\end{equation}
where the second sum includes both orders of each edge.

Fix an ordered pair and index its counter and imbalance by the round $t$.
Choose $a=8\Delta$ if $\sum_t g_t>0$ and $a=0$ otherwise.
Replacing $j_t+g_t$ by its positive part can only reduce its distance to $a$, so
\[
 2(a-j_t)g_t\leq (a-j_t)^2-(a-j_{t+1})^2+g_t^2.
\]
Since $(a-j_1)^2=T$ and $|g_t|\leq1$, summing gives
$\sum_t(a-j_t)g_t\leq T$. Thus replacing $\lambda_{vu}$ by $w_v a/(8\Delta)$
in the round average of~\eqref{eq:round} adds at most $w_v/(8\Delta)$ on this
pair. Its new contribution is $w_v(a_{vu}-x_u(c_{vu}))_+$.
The total increase is at most $\sum_v\deg(v)w_v/(8\Delta)\leq W/8$.
\end{proof}

\subsection{Choosing one power per vertex}
Write $\langle\cdot\rangle$ for the average over all ways of choosing
one retained entry at each vertex.
\begin{lemma}\label{lem:shortage}
For lists of local choices of the same even length $T$, retaining the upper half
at every vertex gives $\langle D_v\rangle\leq 1/2+2e_v$.
\end{lemma}
\begin{proof}
Fix $v$ and put $h=T/2$. Write its $2h$ local choices as rows, each
with its power and $r_v$ requests. Sort the rows by increasing power.
First suppose every neighbor offers at least as often as $v$ requests it.

A neighbor requested at least $h$ times is always active since both
endpoints afford its edge throughout their upper halves. Count these neighbors
first. If they meet the requirement, the shortage is zero. Otherwise
let $d$ be the number still needed. Each row has at least $d$ requests
to the remaining neighbors. Keep any $d$ per row. For each remaining $u$, match its requests
to distinct retained offers at $u$. There are enough since $u$ is
requested fewer than $h$ times.

For each choice of retained entries, mark a request if its assigned
entry is chosen. Thus each neighbor marks at most one request.
Scan the rows in order, taking requests within a row in any order.
Start with a deficit of $d$. At each mark, charge the current deficit
and reduce it by one, down to zero. The positive charges are
$d,d-1,\ldots$, so their sum is at most $d(d+1)/2$.

Fix a request to $u$ and the choices at all other neighbors. Let $Z_i$ be
the preceding deficit for entry $i$ at $u$, with the matched entry first.
Entry~1 marks no earlier request to $u$, so $Z_1\geq Z_i$ for every $i$.
Only entry~1 incurs a charge. The mean charge $Z_1/h$ is therefore at
least $1/h$ times the mean preceding deficit. Averaging over the other
neighbors preserves the inequality.

The first $j$ requests contain $j/h$ marks on average, so the mean
deficit before request $j+1$ is at least $d-j/h$. The $dh$ requests
in the bottom half therefore have mean total charge at least
\[
 \frac1h\sum_{j=0}^{dh-1}\left(d-\frac jh\right)\geq\frac{d^2}{2}.
\]
In an upper row, $v$ affords every edge requested up to that row.
Each mark seen by its end thus gives a distinct active neighbor.
The deficit before each of its $d$ requests is at least that row's
shortage, so the row's mean charge is at least $d/h$ times its mean
shortage. Summing over the $h$ upper rows gives
\[
 d\,\langle D_v\rangle
 \leq \text{mean total charge on the upper rows}
 \leq \frac{d(d+1)}2-\frac{d^2}2=\frac d2.
\]

Finally, if $v$ requests $u$ in $m_u$ rows but $u$ offers in only
$o_u<m_u$, count its largest $m_u$ entries as offers for this $v$ only. Restoring
the true offers can lose one neighbor on at most $m_u-o_u$ retained
entries at $u$. The increase in mean shortage is at most
$\sum_u(m_u-o_u)_+/h=2e_v$.
\end{proof}

\begin{lemma}\label{lem:conditional}
Given a finite list of powers at each vertex and prices $s_v\geq0$,
one can choose one power per vertex with
$\sum_v(p_v+s_vD_v(p))$ at most its product average, in polynomial time.
The same holds for rationally weighted lists and asymmetric endpoint costs.
\end{lemma}
\begin{proof}
Fix one vertex at a time to an entry minimizing the average over the
remaining lists. The old average is a weighted average of the choices.
To compute it, consider $v$ at a given power $b$. For each neighbor
$u$, let $\pi_u$ be the fraction of its list that activates $vu$ at $b$.
The coefficient of $z^i$ in $\prod_{u\in N(v)}(1-\pi_u+\pi_u z)$
is the fraction of choices with $i$ active neighbors. Multiply the
factors, weight the coefficients by $(r_v-i)_+$, and average over
$b$ to obtain the mean shortage at $v$. A fixed vertex has a one-entry
list. The products have polynomial degree and bit length.
\end{proof}

\paragraph{Running time.}
For symmetric costs one scan of the counts at each vertex retains the
largest $h=T/2$ entries, splitting the count at the cutoff if needed.
Suffix sums give the initial offer fractions on all ordered pairs.
These steps take $O(n+|E|)$ operations. Fixing $v$ affects only its own term and those of its
neighbors. The term for a neighbor $u$ takes just two values, according
as $v$ affords $vu$ or not. To compute either value, scan the edges
at $u$ by increasing cost. Insert each factor of the product as its edge
becomes affordable and keep coefficients of degree below $r_u$. This
gives the shortages for all powers at $u$ in $O(\deg(u)(r_u+1))$
operations. The same scan at $v$ gives its own term at each candidate
power. Sum the steps from its neighbors as the power at $v$ increases.
Store the current offer fraction on each ordered pair. Each vertex
$u$ needs $O(\deg(u)+1)$ scans over all fixings. Thus selection takes
$O(n+|E|\Delta^2)$ arithmetic operations. A mean shortage at $u$ has
denominator dividing $h^{\deg(u)+1}$, one factor for $u$ and one per
neighbor. The counters are at most $4\Delta+T$, and the power levels
are input costs or zero. All intermediate numbers have polynomial bit length.

\begin{proof}[Proof of Theorem~\ref{thm:power}]
Every discarded entry at $v$ costs at least $w_v$, so the retained lists
have mean total power at most $2C-W$. Adding the repair bound from
Lemma~\ref{lem:shortage} and using Lemma~\ref{lem:prices} gives
\[
 \langle\Psi\rangle\leq 2C-W+\sum_v w_v(1/2+2e_v)
 =2\Bigl(C+\sum_v w_ve_v\Bigr)-W/2\leq2\OPT-W/4.
\]
Lemma~\ref{lem:conditional} chooses powers within this bound, and
Lemma~\ref{lem:repair} completes the solution. Applying the same repair
bound to the initial powers $w$ gives
$\OPT\leq\Psi(w)\leq W+\sum_v r_v w_v\leq(k+1)W$.
Substitution gives the claimed ratio. Repair scans each adjacency list
at most twice. The $16\Delta^2$ rounds and selection take $O(|V|+|E|\Delta^2\log(\Delta+1))$ arithmetic operations.
\end{proof}

\section{Hardness of approximation}\label{sec:hardness}
Let $H$ be a connected graph on $M\geq2$ vertices with a proper coloring
by nonnegative integers, and let $s$ be its largest color. Replace each vertex of color $i$
by a group of $M^i$ {\bf clones}. Its {\bf high power} is $a_i=M^{-i}$,
so buying high power for a whole group costs one. Call a group high if
every clone in it buys at least $a_i$.
For each edge of $H$ between colors $i<j$, join the two clone groups
by a complete bipartite graph with edge cost $a_j$. A clone $x$ has
requirement $r_x$ equal to its number of higher-colored neighbors.
Call this graph a layer.

Take $k=\max_x r_x$ disjoint layers and add $k$ common {\bf helpers},
each of requirement zero. Join each color-$i$ clone $x$ to the first
$r_x$ helpers at cost $a_i$. Buying all helpers at power one satisfies every high clone and costs
only one per layer on average.

\begin{lemma}\label{lem:gadget}
From a feasible solution of power $C$ one can find in polynomial time a vertex cover of
$H$ of size at most $C/k$. Conversely, a cover of size $q$ gives a
solution of power at most $k(q+2)$. The constructed graph is connected
and simple.
\end{lemma}
\begin{proof}
Consider an edge of $H$ between colors $i<j$. If any clone of its
color-$i$ group buys less than $a_i$, it can use neither a helper nor
a lower-colored neighbor, since those edges cost $a_i$. To meet its requirement it must use every
higher-colored neighbor. Thus all clones of the color-$j$ group must
buy at least $a_j$. The high groups in each layer therefore form a vertex cover. Each such group costs
at least one, so the smallest of the $k$ covers has size at most $C/k$.

Given a cover of size $q$, buy its groups high in every layer and buy
each helper at its largest incident cost. The helpers satisfy every
high clone. Each remaining clone buys the largest cost to a higher-colored
neighbor, or zero if it has none. All its higher-colored neighbors belong
to cover groups, so this satisfies the clone. A color-$i$ clone outside
the cover pays at most $a_i/M$, hence the whole group pays
at most $1/M$. The $k$ layers cost at most $k(q+1)$ and the $k$ helpers
cost at most $k$.

Every positive-requirement clone is adjacent to the first helper.
Each other clone has a lower-colored neighbor of positive requirement,
and each helper is adjacent to a clone of requirement $k$. Thus the graph
is connected. Each pair of vertices receives at most one edge.
\end{proof}

A layer has at most $M^{s+1}$ clones, and $k$ is at most the number of
clones in a layer. Thus for fixed $s$ the construction has polynomial
size. Multiplying all costs by $M^s$ makes them positive integers and
preserves approximation ratios.

\begin{proof}[Proof of Theorem~\ref{thm:lower}]
An edgeless {\sc Vertex Cover} instance has the empty cover. Otherwise
let $\kappa\geq1$ be the minimum cover size of the source graph $G_0$. Choose a fixed integer $\ell\geq2$ so that
$(\ell+3)/(\ell-1)\leq1+\eta$.
Replace each vertex by $\ell$ labeled copies of colors
$0,\ldots,\ell-1$, and replace each source edge by all pairs of copies
of unequal colors. Add a vertex of color $\ell$ adjacent to every copy.
The resulting graph $H$ is connected and its largest color is $s=\ell$.
In the clone construction every lower-colored clone is adjacent to all
$M^\ell$ clones of the new vertex, so $k\geq M^\ell$.

All copies of a minimum cover of $G_0$, together with the new vertex,
form a cover of $H$ of size $\ell\kappa+1$. Conversely, from a cover $Q$ of $H$, select each source
vertex for which at least $\ell-1$ labels belong to $Q$. This gives a
source cover of size at most $|Q|/(\ell-1)$. Indeed, two adjacent
unselected vertices each omit at least two labels. Choosing unequal omitted labels then gives an uncovered edge of $H$.

Apply the clone construction to $H$. Its optimum is at most
$k(\ell\kappa+3)$ by Lemma~\ref{lem:gadget}. A $\gamma$-approximate
solution gives a cover $Q$ of $H$ of size at most $\gamma(\ell\kappa+3)$.
Selecting the vertices of $G_0$ with at least $\ell-1$ copies in $Q$
gives a cover of $G_0$ of size at most
\[
 \gamma\,\frac{\ell\kappa+3}{\ell-1}
 \leq\gamma\,\frac{\ell+3}{\ell-1}\,\kappa
 \leq(1+\eta)\gamma\kappa.\qedhere
\]
\end{proof}

\section{Asymmetric endpoint costs}\label{sec:asymmetric}

\problem{{\sc Activation Edge Multicover}}
{A loopless undirected multigraph $G=(V,E)$, nonnegative rational endpoint costs
$c_e^v,c_e^u$ for every $e=vu$, and integer requirements $r_v\geq0$.}
{Powers $p_v\geq0$ minimizing $\sum_vp_v$, such that every vertex $v$ has
at least $r_v$ distinct neighbors $u$ with some edge $e=vu$ satisfying
$p_v\geq c_e^v$ and $p_u\geq c_e^u$.}

The instance is feasible exactly when $r_v\leq |N(v)|$ at every vertex,
since buying all thresholds activates every pair of neighbors. We assume
this condition throughout.

\subsection{The relaxation and the cost of repair}
Let $B_v=\{0\}\cup\{c_e^v:e\in\delta(v)\}$. At power $b\in B_v$,
the least power needed at $u$ to activate the pair $vu$ is
\[
 q_{vu}(b)=\min\{c_e^u:e=vu,\ c_e^v\leq b\}.
\]
Put $q_{vu}(b)=\infty$ if $v$ cannot afford an edge to $u$ at level $b$.
Then $q_{vu}$ is nonincreasing. Let $w_v$ be the least level at which
$r_v$ neighbors are affordable, with $w_v=0$ if $r_v=0$.

Let $z_{vb}$ be the weight of level $b$ at $v$ and
$x_v(a)=\sum_{b\in B_v:\,b\geq a}z_{vb}$, with $x_v(\infty)=0$.
At level $b$, the variable $y_{vbu}$ is the weight of the request from
$v$ to $u$. It is defined only for affordable neighbors. All variables are
nonnegative. For every threshold $a$ at $u$, the
total weight of requests from $v$ needing at least $a$ must be at most
$x_u(a)$. The LP minimizes $\sum_v\sum_{b\in B_v}b z_{vb}$ subject to
\begin{align}
 \sum_{b\in B_v}z_{vb}&=1
       &&(v\in V),\notag\\
 y_{vbu}&\leq z_{vb}
       &&(v\in V,\ b\in B_v,\ q_{vu}(b)<\infty),\notag\\
 \sum_{u:\,q_{vu}(b)<\infty}y_{vbu}&=r_v z_{vb}
       &&(v\in V,\ b\in B_v),\notag\\
 \sum_{b\in B_v:\,a\leq q_{vu}(b)<\infty}y_{vbu}&\leq x_u(a)
       &&(v\in V,\ u\in N(v),\ a\in B_u).\label{eq:asymmetric-LP}
\end{align}
Every feasible assignment, rounded down to levels in $B_v$, gives an LP
solution by requesting $r_v$ active neighbors at each vertex. A level below $w_v$ has
too few affordable neighbors to carry positive weight. Thus the optimum
$L$ of this polynomial-size LP satisfies $W=\sum_v w_v\leq L\leq\OPT$.

For $r_v>0$ and $b\geq w_v$, let $q_v(b)$ be the $r_v$-th smallest
$q_{vu}(b)$ over neighbors $u$. Define the {\bf repair price}
\[
 s_v=\min_{b\in B_v:\,b\geq w_v}\{b-w_v+q_v(b)\},
 \qquad s_v=0\text{ if }r_v=0.
\]
For $r_v=1$, $w_v+s_v=\min_{e=vu}(c_e^v+c_e^u)$. Thus an edge-cover
instance has slope at most $a$ in the sense of~\cite{NutovKortsarzShalom2019}
exactly when $s_v\leq a w_v$ at every required vertex.

Starting at powers at least $w$, a shortage $D>0$ at $v$ can be
filled at cost at most $D s_v$. Raise $v$ to at least a minimizing level
$b$. If $v$ still lacks neighbors, activate enough of the $r_v$ pairs
with smallest $q_{vu}(b)$ to meet its requirement. The increase is
at most $b-w_v+D q_v(b)\leq D s_v$.
As in Lemma~\ref{lem:repair}, the total repair cost from $p$ is at most
$\sum_v s_vD_v(p)$.

Taking $b=w_v$ gives $s_v\leq\theta w_v$ for finite $\theta$.
Taking $b=p_v^*$ in an optimum assignment gives
$q_v(b)\leq\max_{u\in N(v)}p_u^*$ when $r_v>0$. Since $v$ is distinct
from its neighbors, $s_v\leq p_v^*+\max_{u\in N(v)}p_u^*\leq\OPT$.
Hence $S=\sum_v s_v$ satisfies $S\leq\theta W$ for finite $\theta$ and
\begin{equation}\label{eq:repair-price-sum}
 S\leq\min\{\theta,n\}\,\OPT.
\end{equation}

\subsection{Choosing powers and repairing shortages}

\begin{proposition}\label{prop:asymmetric}
For rational $t\geq4\ln2$, a feasible solution of power at most
\begin{equation}\label{eq:asymmetric-bound}
 W+t(L-W)+S e^{-t/4}
\end{equation}
can be found in polynomial time.
\end{proposition}
\begin{proof}
Take an optimum LP solution. At each vertex retain the upper $1/t$
fraction of its power distribution, splitting the weight at the cutoff
if necessary, and normalize it. The fraction of retained weight at or above $a$ is
$\min\{1,t x_v(a)\}$. The mean total power is at most $W+t(L-W)$,
since every discarded level at $v$ is at least $w_v$.

Fix $v$ and let $b$ be its smallest retained power. Its shortage can
only decrease at higher powers. Write $\pi_u=\min\{1,t x_u(q_{vu}(b))\}$
for the fraction of retained powers at $u$ that activate $vu$ at $b$.
Count the neighbors with $\pi_u=1$
and let $d$ be the remaining requirement. Suppose $d>0$.

Since $t\geq2$, the bottom half of the LP weight at $v$ lies at levels
at most $b$. At a split level divide its requests in the same proportion
as its weight. The bottom half thus allocates $r_v/2$ in total, at most $1/2$ to each
neighbor already counted. Thus the other neighbors receive at least
$d/2$ in total. The amount received by each such $u$ is at most
$x_u(q_{vu}(b))$ by~\eqref{eq:asymmetric-LP}, since the lower levels
require at least $q_{vu}(b)$ at $u$. An unaffordable pair receives zero.
Hence the number $X$ of active neighbors among those remaining has
mean $\mu=\sum_{u:\pi_u<1}\pi_u\geq td/2$.

For every integer $i\geq0$, $(d-i)_+\leq 2^{d-i-1}$. Taking the product average gives
\[
 \langle D_v\rangle\leq 2^{d-1}\langle 2^{-X}\rangle
 =2^{d-1}\!\prod_{u:\pi_u<1}(1-\pi_u/2)
 \leq 2^{d-1}e^{-td/4}\leq e^{-t/4},
\]
where the last inequality uses $t\geq4\ln2$.
Thus the mean power plus the mean repair bound is at most
\eqref{eq:asymmetric-bound}. Lemma~\ref{lem:conditional} applies with
prices $s_v$, since at any fixed $b$ each neighbor is tested against
the single threshold $q_{vu}(b)$. Fix the powers and repair.
\end{proof}

\begin{proof}[Proof of Theorem~\ref{thm:activation}]
Put $m=\min\{\theta,n\}$ and take $t=4\lceil\log_2(2+m)\rceil$.
Then $m e^{-t/4}\leq1$. By~\eqref{eq:repair-price-sum} and
Proposition~\ref{prop:asymmetric}, the total power is at most
$tL+S e^{-t/4}\leq(t+1)\OPT$.
\end{proof}

For $\theta\leq8$, taking $t=4$ in Proposition~\ref{prop:asymmetric}
gives $4L-3W+S/e\leq4L$, since $S\leq8W$.

\paragraph{Use of AI.}
We used large language models to help develop proof strategies, examine
arguments, search for counterexamples, and draft and edit this paper.
We are responsible for its contents.

\bibliography{references}

\end{document}